\documentstyle[prl,aps,epsfig,multicol]{revtex}

\def\inun{i\nu_n}
\def\iomn{i\omega_n}
\def\vq{\vec{q}}
\def\vS{\vec{S}}
\def\vT{\vec{T}}
\def\cS{{\cal M}}
\def\chil{\chi_{\rm loc}}
\def\chinn{\chi_{\rm nn}}
\def\su{s_\uparrow}
\def\tu{t_\uparrow}
\def\sd{s_\downarrow}
\def\td{t_\downarrow}
\def\iomn{i\omega_n}
\def\Sis{\Sigma_s}
\def\Sit{\Sigma_t}

\begin{document}
\draft

\title{Dynamical Mean-Field Theory of Resonating Valence Bond Antiferromagnets}
\author{Antoine Georges, Rahul Siddharthan and Serge Florens}
\address{
Laboratoire de Physique Th\'eorique, Ecole Normale Sup\'erieure,
24 rue Lhomond, 75231 Paris Cedex 05, France \\ and Laboratoire de
Physique des Solides, Universit\'e Paris-Sud, B\^{a}t.~510, 91405
Orsay, France}

\date{Version October 25, 2001; printed \today}
\maketitle

\begin{abstract}
We propose a theory of the spin dynamics of frustrated quantum
antiferromagnets, which is based on an effective action for a
plaquette embedded in a self-consistent bath. This approach,
supplemented by a low-energy projection, is applied to the kagome
antiferromagnet.  We find that a spin-liquid regime extends to very
low energy, in which local correlation functions have a slow decay in
time, well described by a power law behaviour and $\omega/T$ scaling
of the response function: $\chi''(\omega)\propto
\omega^{-\alpha}F(\omega/T)$.
\end{abstract}
\vspace{0.15cm}

\begin{multicols}{2}

The possibility of frustrated quantum antiferromagnets (QAF) having a
resonating valence bond (RVB) ground state, that is, a superposition
of states where all spins are paired into singlets, was suggested many
years ago~\cite{rvb}; such a ground state has neither long-range spin
order (eg Neel order) nor spin-Peierls order (an ordered arrangement
of these singlet pairs). Recently, convincing evidence has been given
that some frustrated two-dimensional QAF indeed have RVB
physics~\cite{lhuil,moessner}. In particular, numerical studies
~\cite{lhuil} of the spin-half Heisenberg QAF on the kagome lattice
reveal that the model has no long-range order and displays a very
small gap to magnetic (triplet) excitations (estimated to be of order
$J/20$).  Moreover, this gap is filled with an exponential number of
singlet excitations, suggesting a possible continuum in the
thermodynamic limit, and a large low temperature entropy, in agreement
with the RVB picture.  We also note, though this is seldom
emphasized, that the results of \cite{lhuil} suggest a correspondingly
large number of triplet states immediately above the gap.  This may be
expected, since in any valence bond component of an RVB state, one can
replace any singlet pair by a triplet pair and continue to have an
eigenstate of $S$.  Resonance between such states would lower the
excitation energy to much below $J$.

These considerations suggest that, in a temperature range above the
triplet gap, the spin correlations have a rapid decay in space, but a
slow decay in time due to the large density of triplet excited states.
Equivalently, the dynamical susceptibility $\chi(\vq,\omega)$ would
not display a narrow peak around a specific ordering wave-vector, but
$\chi''(\vq,\omega)$ would have weight at low frequency,
characteristic of a spin liquid.  Indeed, recent inelastic neutron
scattering studies \cite{mondelli} on the $S=3/2$ kagome slab compound
SCGO reveal such behaviour above the freezing temperature ($T\geq
4\mbox{K}$). The only relevant energy scale in this spin-liquid regime
is apparently set by the temperature itself, with $\chi''_{{\rm
loc}}\equiv\sum_{\vq}\chi''(\vq,\omega)$ obeying a scaling behaviour
$\chi''_{{\rm loc}}\sim \omega^{-\alpha} F(\omega/T)$ with
$\alpha\simeq 0.4$, corresponding to a slow decay in time of the local
dynamical correlations $\langle S(x,0)S(x,t)\rangle\sim
1/t^{1-\alpha}\sim 1/t^{0.6}$. We observe that the uniform
($\vq=\vec{0}$) susceptibility in this regime is well fitted by
$\chi\propto T^{-0.4}$ too.

In this article, we introduce a novel theoretical approach to the spin
dynamics in the spin-liquid regime of frustrated QAF, which focuses on
short-range spin correlations only.  This approach draws inspiration
from the dynamical mean-field theory (DMFT) of itinerant fermion
models \cite{review}, and some of its extensions
\cite{review,cluster-dmft,dca,ext-dmft1,ext-dmft2,qsg}. Our method is
fairly general but is applied here to the concrete case of the
spin-half kagome QAF. The determination of on-site and
nearest-neighbor dynamical spin correlations is mapped onto the
solution of a model of quantum spins on a triangular plaquette coupled
to a self-consistent bath.  (A single-site approach is not adequate,
since the essential physics of singlet formation involves at least
nearest-neighbor sites).  Using a projection onto the low-energy
sector of this model, we demonstrate that, in a temperature range
extending down to $T\ll J$, a slow (power-law) decay of temporal
correlations is found.

\begin{figure}[ht]
\centerline{\epsfig{file=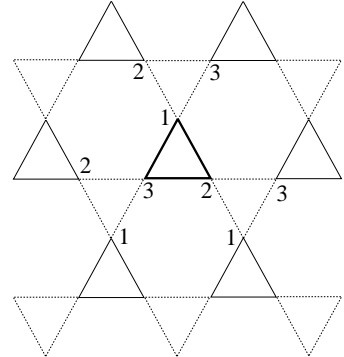,width=4.5cm,clip=}}
\caption{\label{fig_lattice}
 The kagome lattice, viewed as an up-pointing cluster (bold)
embedded in a network of similar clusters, which is approximated
in the DMFT approach by a self-consistent bath. }
\end{figure}

We view the kagome lattice as a triangular superlattice of up-pointing
triangular plaquettes (Fig.~\ref{fig_lattice}).  Sites are labeled by
an index $a=1,2,3$ within a plaquette and by a triangular superlattice
index $I$ numbering the plaquette. We denote by
$\chi_{ab}(\vq,\tau-\tau')$~ the Fourier transform of the dynamical
spin correlation function
$\frac{1}{3}\langle\vec{S}_{a,I}(\tau)\cdot\vec{S}_{b,I'}(\tau')\rangle$
with respect to the plaquette coordinates $I,I'$~ ($\vq$ is a vector
of the supercell Brillouin zone). Our approach relies on approximating
the correlation function by
\begin{equation}
\label{dmft-ansatz}
\chi(\vq,\inun)^{-1}_{ab}=J_{ab}(\vq)+\cS_{ab}(\inun).
\end{equation}
Here, $\nu_n\equiv\,2n\pi/\beta$ is a bosonic Matsubara frequency, and
all quantities are matrices in the internal plaquette indices ($a,b$).
$J_{ab}(\vq)$ is the supercell Fourier transform of the exchange
couplings.  $\cS_{ab}$ is a measure of how much the correlation
function differs from that of a Gaussian model, for which
$\chi=J^{-1}$, and hence plays the role of a spin ``self-energy''
matrix for the QAF. Obviously, the key assumption made in
(\ref{dmft-ansatz}) is that the $\vq$-dependence of this self-energy
can be neglected.  This approximation, which is likely to be
reasonable when spatial correlations are short-ranged, is analogous to
the assumption of a momentum-independent self energy made in the DMFT
of correlated fermion models. Here however, the DMFT concept is
extended on two accounts: the local ansatz is made on the spin
correlation function rather than on a single-fermion quantity, and a
plaquette rather than a single site is considered.  Extensions of DMFT
to clusters \cite{review,cluster-dmft,dca} and to (spin or charge)
response functions \cite{ext-dmft1,ext-dmft2,qsg} have been considered
separately before in different contexts. Combining them is a unique
aspect of our approach, which is necessary to capture the dynamical
aspects of inter-site singlet formation at the heart of spin-liquid
behaviour.

In order to calculate the dynamical self-energy matrix
$\cS_{ab}(\inun)$, an effective action involving only the spins of
a single triangular plaquette is introduced, which reads
\begin{eqnarray}
\nonumber
S & = & S_B + \frac{1}{2}\,J\,\sum_{a\neq
b}\int_0^\beta\!d\tau\, \vec{S}_{a} \cdot\vec{S}_b+\\
 & &+\frac{1}{2}\,\int_{0}^{\beta} \!\!d\tau d\tau'\,
\sum_{ab}\,D_{ab} (\tau -\tau')\,\vec{S}_{a} (\tau ) \cdot
\vec{S}_{b} (\tau'),
\label{action}
\end{eqnarray}
in which $S_B$ denote Berry phase terms. $D_{ab}(\tau-\tau')$ is a
retarded interaction, generated by integrating out all
spins outside the plaquette. Higher order
interactions are also induced in this process, which have been
neglected in (\ref{action}). Equivalently, one can view the rest
of the lattice as an external bath which couples to the spins in
the cluster through time-dependent external fields with a Gaussian
correlator $D_{ab}$. The latter will be determined by a self
consistency requirement, which stipulates that the correlation
functions on a plaquette calculated from the above action
($\chi_{ab}(\tau-\tau')\equiv
\frac{1}{3}\langle\vS_a(\tau)\cdot\vS_b(\tau')\rangle_S$)
involve the same self-energy as the entire lattice. This reads:
$\chi^{-1}_{ab}(\inun)=J\Delta_{ab}+D_{ab}(\inun)+\cS_{ab}(\inun)$,
where $J\Delta_{ab}\equiv\,J(1-\delta_{ab})$ is the matrix of
nearest neighbor couplings on the cluster. Inserting this relation
for the self-energy into (\ref{dmft-ansatz}) and imposing that
$\sum_{\vq}\chi_{ab}(\vq)=\chi_{ab}$ leads to the final form of
the self-consistency requirement
\begin{equation}
\label{scc} \chi_{ab}(\inun)=\sum_{\vq} \,
\left[J(\vq)-J\Delta+\chi^{-1}(\inun)-D(\inun)\right]^{-1}_{ab}.
\end{equation}
We note that the matrix $J(\vq)_{ab}-J\Delta_{ab}$ involves only
the exchange constants linking together different (up-pointing)
triangular plaquettes. In the limit where these inter-plaquette
couplings vanish while the internal ones are kept fixed (decoupled
triangles), our approach becomes exact since (\ref{scc}) implies
$D_{ab}=0$ and (\ref{action}) reduces to the action associated
with the Heisenberg model on a single triangle.
We also note that the above DMFT equations can be derived from a
Baym-Kadanoff formalism in which a functional of the correlation
function and self-energy matrix is introduced in the form
\begin{equation}
\Omega[\chi,\cS]\,=\,\frac{1}{2}\mbox{Tr}\ln[\cS+J] -
\frac{1}{2}\mbox{Tr}[\chi\cS]+\Phi[\chi].
\label{BKfunctional}
\end{equation}
The stationarity conditions $\frac{\delta\Omega}{\delta\chi}=
\frac{\delta\Omega}{\delta\cS}=0$ lead to the above equations when the
exact functional $\Phi$ is approximated by its value for a single
triangular plaquette. The free energy of the model can thus be
calculated in the DMFT approach by inserting the self-consistent
values of $\chi$ and $\cS$ into (\ref{BKfunctional}). Alternatively, a
functional of the correlation function only can be used, along the
lines of Ref.\cite{ext-dmft2}.

In a phase with unbroken translational invariance, the matrix
$\chi_{ab}(\inun)$ in fact reduces to two elements: the local
susceptibility $\chil=\chi_{aa}$ and the nearest-neighbour one
$\chinn=\chi_{ab}\, (a\neq b)$ (and similarly for $D_{ab}$). It is
actually convenient to use the following linear combinations
(proportional to the eigenvalues of the $\chi_{ab}$ and $D_{ab}$
matrices): $\chi_s=3\left(\chi_{\rm loc}+2\,\chi_{\rm nn}\right)\,,\,
\chi_m=3\left(\chi_{\rm loc}-\,\chi_{\rm nn}\right)$,
$D_s=\frac{1}{3}\left(D_{\rm loc}+2\,D_{\rm nn}\right)\,,\,
D_m=\frac{1}{3}\left(D_{\rm loc}-\,D_{\rm nn}\right)$. Introducing the
corresponding self-energies $\cS_s(\inun)\equiv\, 3/\chi_s(\inun)-3
D_s(\inun)\,\,\,,\,\,\,\cS_m(\inun)\equiv\, 3/\chi_m(\inun)-3
D_m(\inun)$, straightforward but tedious algebra allows us to recast
(\ref{scc}) in the form of two scalar equations:
\end{multicols}
\noindent\rule{8.6cm}{0.5pt}
\begin{eqnarray}
\label{scc-scalar}
\frac{\chi_s}{3} & = & \int \frac{\cS_m + J - \frac{2}{3} J\epsilon}
     {\cS_s(\cS_m+J) -2J^2 -\frac{2}{3} (\cS_s-\cS_m)J\epsilon}\,
     \rho(\epsilon) d\epsilon, \nonumber \\
\frac{\chi_m}{3} & = & \int \frac{\cS_s\cS_m - J^2 + \frac{1}{3} J
          (2\cS_m-\cS_s-J)\epsilon}
     {(\cS_m-J)\left( \cS_s(\cS_m+J) -2J^2 -\frac{2}{3}
         (\cS_s-\cS_m)J\epsilon \right)}\,  \rho(\epsilon) d\epsilon.
\end{eqnarray}
\hfill\rule{8.6cm}{0.5pt}
\begin{multicols}{2}\noindent
Here, $\epsilon$ stands for $\cos q_1+\cos q_2+\cos q_3$, where
the $q_i$ are components of $\vq$ along the three directions of
the triangular lattice ($q_1+q_2+q_3=0$), and
$\rho(\epsilon)$ is the corresponding density of states. The
$\vq$-summation in (\ref{scc}) simplifies into an
$\epsilon$-integration in (\ref{scc-scalar}).

Equations (\ref{action},\ref{scc-scalar}) entirely define the
plaquette-DMFT approach introduced here and one could embark at this
stage into a numerical determination of the two key dynamical
quantities $\chil(\tau)\,,\,\chinn(\tau)$ using, e.g., a
generalisation of the quantum Monte Carlo algorithm recently
introduced in \cite{marcelo} in the context of quantum spin-glasses.
Instead, we shall gain further insight into the problem by making use
of a projection onto the low-energy sector of the Hilbert
space~\cite{subr}.  Interesting insights into the low-energy excited
states of the singlet sector have been recently obtained by Mila and
Mambrini starting from the limit of decoupled up-pointing
triangles~\cite{mila1,mila2}.  Each such triangle has a four-fold
degenerate ground-state with spin $S=1/2$ and a four-fold degenerate
excited state with spin $S=3/2$. Hence the lattice of decoupled
triangles has an exponentially large number of ground-states, equal to
$4^{N_s/3}$. The physical picture of \cite{mila1,mila2} is that the
degenerate ground state broadens into a band as the inter-triangle
coupling is turned on. In the following, we shall project the DMFT
equations onto the low-energy Hilbert space corresponding to the
$S=1/2$ sector of each triangle. This amounts to first diagonalizing
the single-plaquette hamiltonian (neglecting $D_{ab}$) and then
rewriting the retarded term in the action (\ref{action}) in this
low-energy sector. To this end, it is convenient to follow
\cite{subr,mila1} and choose as a basis of the 4-fold degenerate
ground-state the eigenvectors of the two following operators: the
spin-$1/2$ operator corresponding to the total spin $\vS$ in a
plaquette, and a doublet of Pauli matrices $\vT=(T^x,T^y)$ with the
meaning of a plaquette chirality operator. In terms of those, the
low-energy projection of the three spin operators in a plaquette read
(with $\omega\equiv\exp{i2\pi/3}$ and for $a$ $=$ 1,2,3)
\begin{equation}
\vS_a^{\rm low} = \frac{1}{3}\vS\,\left[1-2\omega^{1-a}T^- -
  2\omega^{a-1}T^+ \right]
\end{equation}
The low-energy projection of the action (\ref{action}) is then
easily obtained as
\begin{eqnarray}\label{action-low}\nonumber
S_{\text{low}} & = & S_{B} +  \int_{0}^{\beta} \!\!d\tau d\tau'\,
\big[\frac{1}{2}D_s(\tau -\tau') +
\\& & + D_m(\tau-\tau')\,\vT(\tau)\cdot\vT(\tau')\big]\,\,
\vS(\tau)\cdot\vS(\tau'),
\end{eqnarray}
where $D_s$ and $D_m$ are as defined earlier. It turns out that
$\chi_s$ and $\chi_m$ have simple interpretations as the ``total
spin'' and ``mixed'' correlation functions:
\begin{eqnarray}\nonumber
\chi_s(\tau-\tau')&\equiv&\frac{1}{3}\langle\vS(\tau)\cdot\vS(\tau')\rangle\,\\
\chi_m(\tau-\tau')&\equiv&\frac{1}{3}\langle\vS(\tau)\cdot\vS(\tau')\,
\vT(\tau)\cdot\vT(\tau')\rangle\,
\end{eqnarray}

To solve the local quantum problem defined by~(\ref{action-low}), we
use an approximate technique that has proven successful in recent studies
of quantum spin-glass models \cite{qsg} and of impurities in
quantum antiferromagnets \cite{spin-qaf}. We introduce two
doublets of spin-1/2 fermions $(\su,\sd)$ and $(\tu,\td)$ subject
to the constraints: $\su^+\su+\sd^+\sd=\tu^+\tu+\td^+\td=1$ in
order to represent the operators $\vS$ and $\vT$ as:
$S^+=\su^{+}\sd\,\,,S^-=\sd^+\su\,\,,S^z=(\su^+\su-\sd^+\sd)/2\,\,;
T^+=\tu^{+}\td\,\,,T^-=\td^+\tu$. The interacting fermion problem
corresponding to (\ref{action-low}) is then solved in the
self-consistent Hartree approximation. Also, the local constraint
is approximated by its average (the associated Lagrange multiplier
is zero due to particle-hole symmetry). Introducing the Green's
functions $G_s$ and $G_t$ for the two fermion fields and defining
self-energies by $G_s^{-1}(\iomn)=\iomn-\Sis(\iomn)\,\,,\,\,
G_t^{-1}(\iomn)=\iomn-\Sit(\iomn)$ (with
$\omega_n=(2n+1)\pi/\beta$ a fermionic Matsubara frequency), this
yields the imaginary-time equations:
\begin{eqnarray}\nonumber
\Sis(\tau) &=& -\frac{3}{8}D_s(\tau)G_s(\tau)+
                 3D_m(\tau)G_s(\tau)G_t(\tau)G_t(-\tau),\\
\Sit(\tau) &=& 3D_m(\tau)G_t(\tau)G_s(\tau)G_s(-\tau).
\label{hartree}
\end{eqnarray}
Then the local problem (for a given bath
$(D_s,D_m)$) reduces to two coupled non-linear integral equations.
The correlation functions are given by
\begin{eqnarray}\nonumber
\chi_s(\tau) &=& -\frac{1}{2}\,G_s(\tau)G_s(-\tau),\\
\chi_m(\tau) &=& 2\,G_s(\tau)G_s(-\tau)G_t(\tau)G_t(-\tau).
\label{chi-green}
\end{eqnarray}
In practice, we use the following algorithm: we start with an
initial guess for the bath $D_{s,m}$, and obtain the Green's
functions by iteration of eq.~(\ref{hartree}).  We can then
calculate the susceptibilities from~(\ref{chi-green}), as well
as the self energies $\cS_{s,m}$.  Inserting the latter
into~(\ref{scc-scalar}) yields new $\chi_{s,m}$ and, in turn,
the new baths $D_{s,m}^{\rm new}(\inun)=-\cS_{s,m}(\inun)/3 +
1/\chi_{s,m}(\inun)$.

\begin{figure}[ht]
\centerline{\epsfig{file=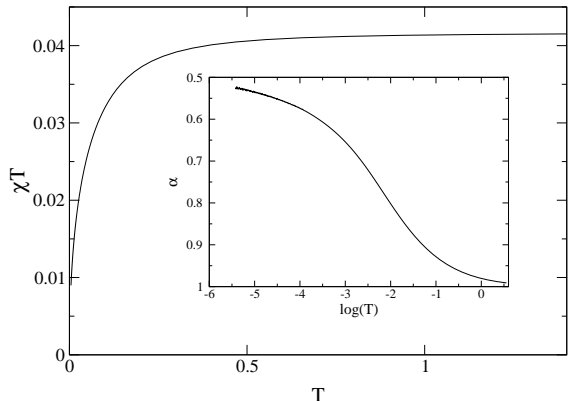,width=7.5cm,clip=}}
\caption{\label{chiloc} {\it Main plot:}
 $\chi_{\rm loc}$ obeys Curie law ($\propto 1/T$)
 at high temperature, but
 diverges more slowly than $1/T$ at lower temperature.
 {\it Inset:} Evolution of the effective exponent $\alpha$
 from high to low temperature
 when fitting $\chi_{\rm loc}(T)$ to $1/T^\alpha$.  ($T$ is in
 units of $J$ in all figures.)
}
\end{figure}

We now discuss our findings when solving these coupled equations
numerically. The first key observation is that a paramagnetic solution
can be stabilized down to the lowest temperature we could reach, with
no sign of long-range ordering intervening. Long-range order is
associated with a diverging eigenvalue of $\chi_{ab}(\vq)$ for some
$\vq$ and hence, within our scheme, to a vanishing denominator in
(\ref{scc-scalar}), which we never observe. In Fig.~\ref{chiloc}, we
display our results for the temperature dependence of the local (i.e
on-site, or $\vec{q}$-integrated) susceptibility $\chi_{\rm loc}(T)$
$=\int_0^\beta \chi_{\rm loc}(\tau) d\tau$. At high temperatures
$\chi_{\rm loc}\propto 1/T$ obeys Curie's law.  Below $T \approx
0.5J$, the effective Curie constant decreases with decreasing $T$,
indicating gradual quenching of the local moment.  However, $\chi_{\rm
loc}$ itself continues to increase as the temperature is lowered.  In
the inset of Fig.~\ref{chiloc}, we display the effective exponent
defined by $\alpha(T)\equiv -\ln\chi_{\rm loc}/\ln T$. A transient
regime, corresponding to a slowly decreasing $\alpha$, is apparent and
extends over more than two decades.  At the lowest temperatures,
$\alpha$ appears to saturate to a value $\alpha\simeq 0.5$.  

\begin{figure}[ht]
\centerline{\epsfig{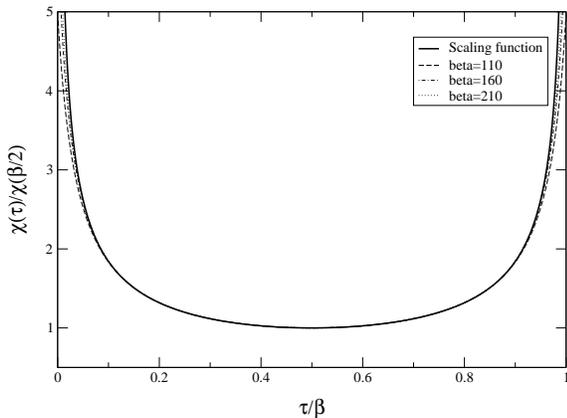}}
\caption{\label{scaledplot}
 For different (low) temperatures, $\chi_{\rm loc}(\tau)/\chi_{\rm
loc}(\beta/2)$ plotted as
 a function of $\tau/\beta$ collapse on a single curve, well
 described by $[\sin(\pi\tau/\beta)]^{-0.5}$ (solid curve). }
\end{figure}

Correspondingly, the correlation function $\chi_{\rm loc}(\tau)$ obeys
a power-law dependence on time in the low-temperature regime.
Figure~\ref{scaledplot} shows $\chi_{\rm loc}(\tau)/\chi_{\rm
loc}(\beta/2)$ as a function of $\tau/\beta$ for several (large)
values of $\beta=1/T$; the graphs collapse on each other for different
$\beta$, and are well fitted by the expression $\chi_{\rm
loc}(\tau)/\chi_{\rm loc}(\beta/2)= \left[\sin\,
\pi\tau/\beta\right]^{-(1-\alpha)}$, with $1-\alpha\simeq 0.5$. This
scaling function is the one appearing in the study of quantum
spin-glasses \cite{qsg} and quantum impurity models
\cite{spin-qaf,kondo} and dictated in the latter case by conformal
invariance. It corresponds to the following scaling form of the
dynamical susceptibility: $\chi''_{\rm loc}(\omega)\propto
\omega^{-\alpha} F_{\alpha}(\omega/T)$ with $F_\alpha(x) = x^\alpha
|\Gamma(\frac{1-\alpha}{2}+i\frac{x}{2\pi})|^2\sinh \frac{x}{2}$.
This also implies a diverging behaviour of the relaxation rate
$1/T_1\propto T\chi''_{\rm loc}(\omega)/\omega\sim 1/T^{\alpha}$.

These findings can be interpreted as the formation of a spin-liquid
regime with a large density of triplet excited states, consistent with
ref.\ \cite{lhuil}.  Our approach fails  to predict a triplet gap,
however, because our triangular plaquette has a spin-1/2 ground-state,
and the classical bath to which it is coupled cannot fully screen this
local moment. Hence the local susceptibility continues to increase at
low temperatures in our approach, down to an unphysically low energy
scale (of order $J/100$). In fact the triplet gap for the $S=1/2$
kagome QAF is known \cite{lhuil} to be quite small (of order $J/20$),
so our description of the spin dynamics should be valid down to a rather 
low temperature scale.  The power-law behaviour of the spin correlations
and the low-temperature increase of the susceptibility agree well with
experimental findings on SCGO above its freezing temperature at $T=4
{\rm K}$ \cite{mondelli}.  Since this is a $S=3/2$ system, it is
conceivable that higher spin extends the range of applicability of our
approach even further. A promising application of our approach is the
pyrochlore lattice: the natural plaquette is a tetrahedron, with a
twofold degenerate singlet ground state, so that the very low-energy
singlet sector should be accessible within our approach.  In future
work we plan to address these issues, and also study the
low-temperature thermodynamics by solving the self-consistent local
problem (\ref{action}) using exact numerical techniques. 

We are grateful to O. Parcollet for his very useful help with
numerical routines, and to C.~Lhuillier, G.~Misguich, R.~Moessner and
S.~Sachdev for discussions. We also thank C.~Mondelli for sharing her
data with us.

\end{multicols}

\begin{references}

\bibitem{rvb} P.W.~Anderson, { Mater. Res. Bull.} {\bf 8}, 153
(1973).
\bibitem{lhuil} C. Waldtmann, H.-U. Everts, B. Bernu, C. Lhuillier, P.
Sindzingre, P. Lecheminant, L. Pierre, { Eur.\ Phys.\ J. B}
{\bf 2}, 501 (1998); P. Lecheminant, B. Bernu, C. Lhuillier, L. Pierre and
P. Sindzingre, Phys. Rev. B {\bf 56}, 2521 (1997).
\bibitem{moessner} R. Moessner and S. L. Sondhi, Phys.\ Rev.\ Lett.\
  {\bf 86}, 1881 (2001); R.~Moessner, S.~L.~Sondhi, E.~Fradkin,
   cond-mat/0103396.
\bibitem{mondelli} C.~Mondelli, H. Mutka, C. Payen, B. Frick, K.H.
Andersen, {Physica B} {\bf
284}, 1371 (2000); C.~Mondelli, { PhD Thesis} (Grenoble, 2000,
unpublished).
\bibitem{review} A.~Georges, G.~Kotliar, W.~Krauth, and M.~Rozenberg,
Rev.\ Mod.\ Phys.\ {\bf 68}, 13 (1996).
\bibitem{cluster-dmft} A.~Schiller and K.~Ingersent, { Phys.
Rev. Lett.} {\bf 75}, 113 (1995)
\bibitem{dca} M. H.~Hettler, A. N. Tahvildar-Zadeh, M. Jarrell,
  T. Pruschke, H. R. Krishnamurthy, {Phys. Rev. B} {\bf
58}, 7475 (1998).
\bibitem{ext-dmft1} J.~Lleweilun Smith and Q.~Si, Phys.\ Rev.\ B
{\bf 61}, 5184 (2000).
\bibitem{ext-dmft2}R.~Chitra and G.~Kotliar, Phys.\ Rev.\ B
{\bf 63}, 115110 (2001).
\bibitem{qsg} A.~Georges, O.~Parcollet and S.~Sachdev, {\sl Phys.
Rev. Lett} {\bf 85}, 840 (2000); {Phys. Rev. B} {\bf 63}.
134406 (2001).
\bibitem{marcelo} D.R.~Grempel and M.J.~Rozenberg {\sl Phys. Rev.
Lett.} {\bf 80}, 389 (1998).
\bibitem{subr} V.~Subrahmanyam, Phys.\ Rev.\ B {\bf 52}, 1133 (1995)
\bibitem{mila1} F.~Mila, Phys.\ Rev.\ Lett.\ {\bf 81}, 2356 (1998)
\bibitem{mila2} M.~Mambrini and F.~Mila, {Eur. Phys. J. B}
{\bf 17}, 651 (2000).
\bibitem{spin-qaf} M.~Vojta, C.~Buragohain and S.~Sachdev, {
Phys. Rev. B} {\bf 61}, 15152 (2000).
\bibitem{hilbert} In fact, a useful approximation to eq.~(\ref{scc-scalar})
is to replace the Hilbert transform of $\rho(\epsilon)$ by its
high-frequency behaviour.
\bibitem{kondo} O.~Parcollet, A.~Georges, G.~Kotliar and A.~Sengupta, 
{ Phys. Rev. B.} {\bf 58} 3794 (1998).

\end{references}
\end{document}